\documentclass[aps,prx,twocolumn,amsmath,amssymb,notitlepage,noeprint,nofootinbib,superscriptaddress]{revtex4-1}
\pdfoutput=1

\usepackage{lineno}
\usepackage[english]{babel}
\usepackage{color}
\usepackage{graphicx}
\usepackage{tabularx}
\usepackage[caption=false]{subfig} 
\usepackage{amsmath,amssymb,bm}
\usepackage[version=3]{mhchem}
\usepackage{verbatim}
\usepackage{multirow}
\usepackage{dcolumn}
\usepackage{float}
\usepackage{nicefrac}
\usepackage{siunitx}
\usepackage{booktabs}
\DeclareSIUnit[number-unit-product = {\,}]{\au}{a.u.}
\DeclareSIUnit[number-unit-product = {\,}]{\kJmol}{\kilo\joule\per\mol}
\definecolor{darkblue}{rgb}{0,0,0.6}
\usepackage[colorlinks,linkcolor=darkblue,citecolor=darkblue,urlcolor=darkblue]{hyperref}

\definecolor{orange}{rgb}{0.93, 0.53, 0.18}
\definecolor{bostonuniversityred}{rgb}{0.8, 0.0, 0.0}
\definecolor{darkspringgreen}{rgb}{0.09, 0.45, 0.27}
\newcommand{\cs}[1]{\textcolor{black}{#1}}

\usepackage{setspace}
\usepackage{subfiles}
\begin{document}
%\linenumbers

% \doublespacing
\title{%
Defects induce phase transition from dynamic to static rippling in graphene
}

\author{Fabian L. Thiemann}
\email{fabian.thiemann@ibm.com}
\affiliation{%
IBM Research Europe, Keckwick Lane, Daresbury, WA4 4AD, United Kingdom
}%
\affiliation{%
Yusuf Hamied Department of Chemistry, University of Cambridge, Lensfield Road, Cambridge, CB2 1EW, United Kingdom
}
\author{Camille Scalliet}
\email{camille.scalliet@ens.fr}
\affiliation{Laboratoire de Physique de l'Ecole Normale Sup\'erieure, ENS, Universit\'e PSL, CNRS, Sorbonne Universit\'e, Universit\'e de Paris, F-75005 Paris, France}
\author{Erich A. Müller}
\affiliation{%
Department of Chemical Engineering, Sargent Centre for Process Systems Engineering, Imperial College London,
South Kensington Campus, London SW7 2AZ, United Kingdom
}%
\author{Angelos Michaelides}
\email{am452@cam.ac.uk}
\affiliation{%
Yusuf Hamied Department of Chemistry, University of Cambridge, Lensfield Road, Cambridge, CB2 1EW, United Kingdom
}

\date{\today}

\begin{abstract}
Two-dimensional (2D) materials display nanoscale dynamic ripples that significantly impact their properties.
Defects within the crystal lattice are the elementary building blocks to tailor the material's morphology.
\cs{While some studies have explored the link between defective structures and rippling dynamics in 2D materials, a comprehensive understanding of this relationship has yet to be achieved.}
\cs{Here, we address this using machine learning-driven molecular dynamics simulations.}
Specifically, we find that above a critical concentration of defects, free-standing graphene sheets undergo a dynamic transition from freely propagating to static ripples.
Our computational approach captures the dynamics with atomic resolution, \cs{and reveals that the transition is driven by elastic interactions between defects.}
\cs{The strength of these interactions is found to vary across defect types and we identify a unifying set of principles driving the dynamic-to-static transition in 2D materials.}
Our work not only rationalises puzzling experimental results for defective 2D materials,  but also paves the way to design two-dimensional devices with tailored rippling dynamics.
These insights could lay the foundations for a new class of disorder-based catalytic and interfacial materials.
\end{abstract}

\maketitle

\section{Introduction}

Although commonly described as flat, two-dimensional (2D) materials owe their existence \cite{Mermin1968} as well as many of their unique properties \cite{geim_rise_2007,lee_measurement_2008,neto_electronic_2009,balandin_thermal_2011} to the presence of intrinsic ripples at the nanoscale \cite{Meyer2007, Fasolino2007, MoS2, hBN}.
For instance, these dynamic out-of-plane fluctuations are known to play a crucial role in the emergence of a diversity of physical phenomena, such as electron-hole puddle formation \cite{martin_observation_2008, Guinea2008a}, suppression of weak localisation \cite{morozov_strong_2006}, enhanced chemical reactivity \cite{elias_control_2009,Boukhvalov2009,sun_unexpected_2023}, and the motion of adsorbates \cite{ma_fast_2016,marbach_transport_2018} across 2D surfaces. 
These observations have triggered significant interest in controlling the structure~\cite{lui_ultraflat_2009, Bao2009, Lehtinen2013, Ludacka2018, Thiemann2020, Singh2022} and dynamics \cite{dynamic_ripples,xu2014unusual, Ackerman2016, Hu2016, lin2017ultrafast} of the ripples to achieve desired material properties.
The artificial insertion of \cs{atomic} defects via electron beam \cite{Krasheninnikov2007,Robertson2012,Susi2017,Zhao2017}
has become the preferred method to precisely manipulate the crystal lattice of 2D materials.
However, understanding how material properties, such as rippling dynamics, emerge from their atomic structure remains a central problem in statistical physics and materials science.\\

This understanding could provide a framework for reconciling puzzling or conflicting experimental results on thermal \cite{Lopez-Polin2017} and mechanical \cite{Zandiatashbar2014,Nicholl2015,Lopez-Polin2015,Lopez-Polin2022} properties of defective graphene.
Controlling the rippling dynamics via defect engineering could open new avenues to develop innovative nano devices. This is particularly motivated by recent reports that `trampoline' dynamics prevents proteins from unfolding \cite{anggara_landing_2023}, and findings that ripples enable graphene to split molecular hydrogen orders of magnitude faster than the best catalysts \cite{sun_unexpected_2023}. \cs{Another motivation arises from the field of nanofluidics, where two-dimensional materials play a pivotal role in confining fluids at the nanoscale. At such scales, the rippling dynamics of interfaces significantly influence fluid flows, with studies predicting rippling-accelerated molecular transport~\cite{ma_fast_2016,marbach_transport_2018}. Understanding the effect of defects on these rippling dynamics could unlock new opportunities to control nanofluidic transport via defect engineering. This insight could, for example, inspire the design of advanced two-dimensional membranes capable of dynamically separating similar molecular species, drawing on concepts demonstrated in bulk porous materials~\cite{su2022separating}.\\} 

Yet, experimental investigations of rippling dynamics have been restricted to defect-free 2D materials. State-of-the-art methods, such as scanning tunnelling microscopy (STM) \cite{xu2014unusual, neek-amal_thermal_2014, Ackerman2016} and ultrafast electron diffraction (UED) \cite{Hu2016, lin2017ultrafast}, achieve atomic and femtosecond resolution.
Still, STM is not well suited to simultaneously track the vertical motion of large regions, and the interpretation of UED-derived diffraction patterns becomes complex in the presence of disorder~\cite{UEDreview}. 
Therefore, it remains unclear whether these techniques can capture the spatial heterogeneity inherent to defective materials.
Similarly, theoretical work becomes exceedingly challenging when moving beyond the study of static properties or ideal crystalline materials. 
Conversely, molecular simulations can in principle provide the required resolution of the rippling dynamics but are highly sensitive to the underlying description of the atomic interactions.
In contrast to simple classical force fields, first-principle approaches accurately describe defect formation energies and phonon dispersion curves.
Still, their computational cost makes them prohibitive for resolving length and time scales beyond a few nanometers and hundreds of picoseconds.\\

Here, we employ large scale molecular dynamics (MD) simulations to explore the impact of atomic defects on the rippling dynamics of free-standing graphene, the prototype 2D material. 
\cs{Using the accurate GAP-20 machine learning potential for carbon \cite{Rowe2020b} }allows us to resolve the dynamics with first principles accuracy, over large length and time scales inaccessible to expensive \textit{ab initio} methods \cite{Behler2016, Deringer2019}.
This is necessary to account for spatially and temporally extended ripples \cite{Fasolino2007}.
We reveal the existence of a defect-induced dynamical transition from freely propagating ripples to frozen and static buckling \cs{upon increasing defect concentration.
By achieving atomic resolution, we reveal that the transition originates from elastic interactions between defects. We show that strongly interacting defects give rise to a sharp dynamical transition, while weakly interacting defects yield a smooth crossover in the dynamics of the defective sheet.}
This knowledge provides guiding principles to tune dynamic rippling, offering prospects for designing tailor-made nanodevices.

\begin{figure*}[t!]
    \centering
    \includegraphics[width=\textwidth]{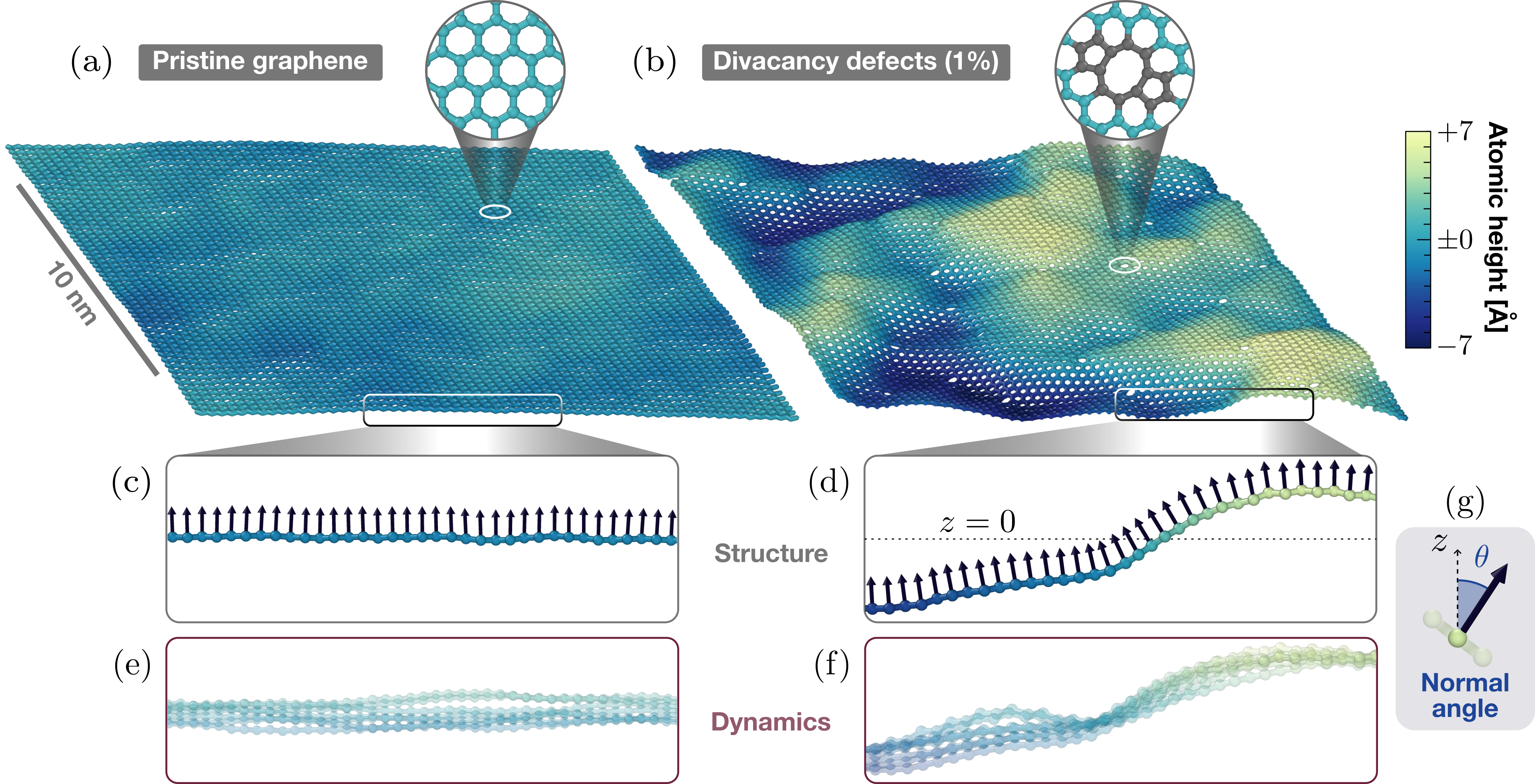}
    \caption{
    {\bf Impact of atomic defects on the structure and dynamics of free-standing graphene.}
    The atoms in (a) pristine and (b) defective free-standing graphene are coloured according to their out-of-plane position relative to the center of mass of the respective system.
    The latter contains 1\% divacancy defects (inset). \cs{See figure S1 in the SI for 1\% Stone-Wales defects.}
    We characterize the structure (c, d) and dynamics (e, f) of graphene via the local inclination $\theta$, defined in panel (g) as the angle between the normal vector (arrows) and the $z$ direction. 
    The side-view cuts of (c, e) pristine and (d, f) defective samples are shown with the same field of view.
    Defects have a profound influence not only on the static corrugation of graphene, but also on its dynamic fluctuations.
    This is illustrated in panels (e, f) where snapshots captured at consecutive times, separated by 1 ps, are superimposed.
    While the pristine sheet evolves freely in (e), dynamic fluctuations are highly constrained by the presence of defects (f).
    \label{fig:overview}}
\end{figure*}

\section{Results}

\begin{figure*}[t!]
\includegraphics[width=\textwidth]{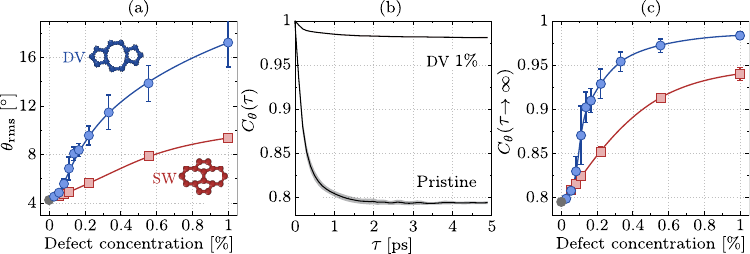}
\caption{{\bf Defect-induced transition in the rippling dynamics of graphene.} (a) Evolution of the root-mean-square inclination $\theta_{\mathrm{rms}}$ with increasing concentration of divacancy (DV, blue) \cs{and Stone-Wales (SW, red) defects}. (b) Normalised inclination time-correlation function $C_\theta(t)$ in pristine and 1\% divacancy defects. (c) Long-time inclination correlation as a function of DV (blue) \cs{and SW (red) defect concentration}. \cs{The error bars represent the standard error of the mean over different defect arrangements.}
}
\label{fig:quant_results}
\end{figure*}

\cs{Among the various atomic defects identified in graphene~\cite{Yang2018}, our focus lies on two saturated defects, namely divacancy (DV) and topological Stone-Wales (SW) defects. 
The former results from the removal of two adjacent carbon atoms induced for example via electron or ion beam irradiation \cite{Krasheninnikov2007,Robertson2012,Susi2017,Zhao2017}. 
The reconstruction of the carbon sp$^2$ network creates an atomic defect comprising an eight-membered ring surrounded by two five-membered rings, as depicted in the inset of Fig. \ref{fig:overview}(b). The latter is obtained by the $90^{\circ}$ rotation of a C-C bond, illustrated in the bottom right of Fig.~\ref{fig:quant_results}(a). 
We choose these two specific defects because their impact on the structure of graphene is already established: divacancies induce strong out-of-plane deviations, while Stone-Wales defects have a smaller effect on corrugation~\cite{Thiemann2021}. 
Here, we investigate the effect of these defects on the rippling dynamics of free-standing graphene. 
Although monovacancies are common in graphene, we do not investigate them because the induced spin polarisation and magnetic moment can only be accurately captured through more accurate electronic structure methods~\cite{monovacancies2016}. 
The comparison between two distinct defects allows us to discuss the generalisability of our results to other two-dimensional materials such as hexagonal boron nitride (hBN) or transition-metal dichalcogenide (TMDC) monolayers, and other types of atomic defects in the Discussion section.} 

Starting with general observations, we present in Fig.~\ref{fig:overview} snapshots of pristine (a,c,e) and defective graphene containing 1\% divacancies (b,d,f). \cs{The corresponding figure with 1\% Stone-Wales defects is provided in the Supplementary Information (SI).}
Freely suspended graphene \cite{Meyer2007} at room temperature is naturally corrugated, with atomic out-of-plane positions deviating from the center of mass position $z=0$, where $z$ is the direction perpendicular to the sheet. 
In the presence of defects, however, the corrugation is significantly greater compared to that induced by thermal fluctuations, as shown in Fig.~\ref{fig:overview}(a-d), which uses the same color scale for both pristine and defective graphene.
Beyond the deviations in the static structure, we observe a critical influence of defects on the dynamics of graphene as evident from Fig.~\ref{fig:overview}(e, f) where we show a superposition of side-views of snapshots at successive times. 
Corresponding movies are provided in the Supplementary Information (SI).
Individual snapshots are slightly transparent, such that dynamic regions of the samples appear blurred, while static regions are darker. 
Pristine graphene displays high flexibility allowing each atom to explore the full range of accessible out-of-plane ($z$) positions, indicating random and short-lived height fluctuations. 
In stark contrast, the highly defective system (right) exhibits dynamic fluctuations confined around a static corrugated pattern. \\

\cs{To comprehend these differences, we now quantify such rippling dynamics; with the key results summarised in Fig.~\ref{fig:quant_results}.
Our approach is based on assessing the local inclination $\theta$, defined as the angle between the surface normal and $z$, see Fig.~\ref{fig:overview}(g). Details on its computation are provided in the Methods section.
At any given time, the root mean square (rms) inclination $\theta_{\mathrm{rms}}$ characterizes how corrugated the sheet is.
We present in Fig.~\ref{fig:quant_results}(a) the dependence of the rms inclination with defect concentration, for DV (blue) and SW (red) defects. 
The pristine value (grey point) serves as a benchmark against transmission electron microscopy (TEM) diffraction experiments, as we find $\theta_{\mathrm{rms}}\approx 4.2^{\circ}$ which agrees well with the experimental value $\theta^\mathrm{exp}_\mathrm{rms}\approx 5.0^{\circ}$.
While experimental reports of the rms inclination are limited to pristine graphene, here we report its dependence on defect concentration.
At each concentration, the data reported corresponds to an average over several defect arrangements (distances and orientations). Individual data points are reported in the SI in figure S19.
We find that DV defects have a strong impact on corrugation, with a fourfold increase from pristine to 1\% DV. The impact of SW defects, on the other hand, is milder as we measure $\theta_{\mathrm{rms}}\approx 10^{\circ}$ at 1\% SW.
These observations align with our prior computational study \cite{Thiemann2021}.
Yet, in contrast with the previously reported steady increase in corrugation with DV concentration, the rms inclination exhibits a sharper increase around 0.1\% divacancies. 
We reveal in Fig.~S12 (SI) that the jump gets sharper with increasing system size.}\\

Next, we employ the local inclination to investigate the influence of defects on the dynamical behavior.
In Fig.~\ref{fig:quant_results}(b) we show the normalised autocorrelation functions of the inclination, $C_\theta (\tau)$, for a pristine sample and graphene \cs{with $1$\% DV defects}.  
A comprehensive overview of all systems studied \cs{(DV and SW defects at various concentrations)} is provided in section S2.A of the SI.
For both systems, $C_\theta (\tau)$ decays exponentially to distinct plateau values within a few picoseconds. \cs{We verify in section S3.A.3 in the SI that the simulation length is sufficient to ensure the convergence of our results and adequately resolve the relevant frequencies.}
The plateau value, that we call long-time inclination correlation and denote $C_\theta(\tau \to \infty)$,
quantifies how much atomic inclinations are correlated over time. 
Values close to 1 suggest a constrained local inclination and static rippling, while lower values indicate greater flexibility and dynamic rippling. \cs{We observe a clear difference between the inclination dynamics of pristine and defective samples. For pristine graphene at room temperature, the fluctuations of the inclination yield a reference value $C_\theta(\tau \to \infty) \approx 0.8$. Instead, the long-time inclination correlation of the 1\% DV samples is very close to one indicating a constrained dynamics, as anticipated from the stroboscopic visualization of Fig.~\ref{fig:overview}(f).} \\

Our analysis thus focuses on comparing the long-time inclination correlation across defect types and concentrations.
\cs{We show the results for DV (blue) and SW (red) defects in Fig.~2(c). Here again, data points are averages over different defect arrangements. Details are provided in the Methods section and in section S2.B of the SI.}
\cs{The lowest value is reached for pristine graphene (grey point). 
The long-time inclination correlation increases as soon as defects are introduced, regardless of their type (DV or SW). This implies that defects partially freeze the rippling dynamics of the sheet.
At low defect concentration, the growth is linear with the number of defects. Interestingly, the same trend describes quantitatively the DV and SW data. 
Yet, the data departs from a linear growth at a concentration that depends on the defect: 0.055\% for DV (from the third data point), while the SW data is linear up to four times larger concentrations, \textit{i.e.} 0.22\% (fourth point).} \\

\cs{Beyond the linear regime, the two situations differ. We observe a sharp increase in the long-time inclination correlation around a DV concentration of $\approx 0.1$\%. This sharp increase mirrors that observed in the rms inclination Fig.~\ref{fig:quant_results}(a), but is more striking in the dynamics. The DV defects thus have a much stronger impact on rippling dynamics than on corrugation. Concomitant with this sharp increase, the variance across different spatial realisations (captured by the error bars) is maximal at 0.1\% DV.}
Above concentrations of $0.15$\%, the plateau values increase steadily again with defect concentration, and seems to have converged to its large concentration value at 1\% DV.
These results suggest a previously unreported disorder-induced dynamical transition from random thermal fluctuations ($C_\theta(\tau \to \infty) \approx 0.8$) to frozen ripples ($C_\theta(\tau \to \infty) \lessapprox 1$) \cs{when divacancy defects are introduced in graphene}.
This conclusion is supported by our results reported in section S4.A.1 of the SI which demonstrate that the transition becomes sharper with increasing system size, suggesting a genuine dynamical transition in the macroscopic limit.
\cs{The case of SW defects is qualitatively different as the inclination dynamics shows no abrupt change upon increasing their concentration. Instead, the long-time inclination grows smoothly from the linear regime towards a high-concentration plateau value, which is smaller than for DV defects. }\\

\begin{figure*}[t!]
    \centering
    \includegraphics[width=\textwidth]{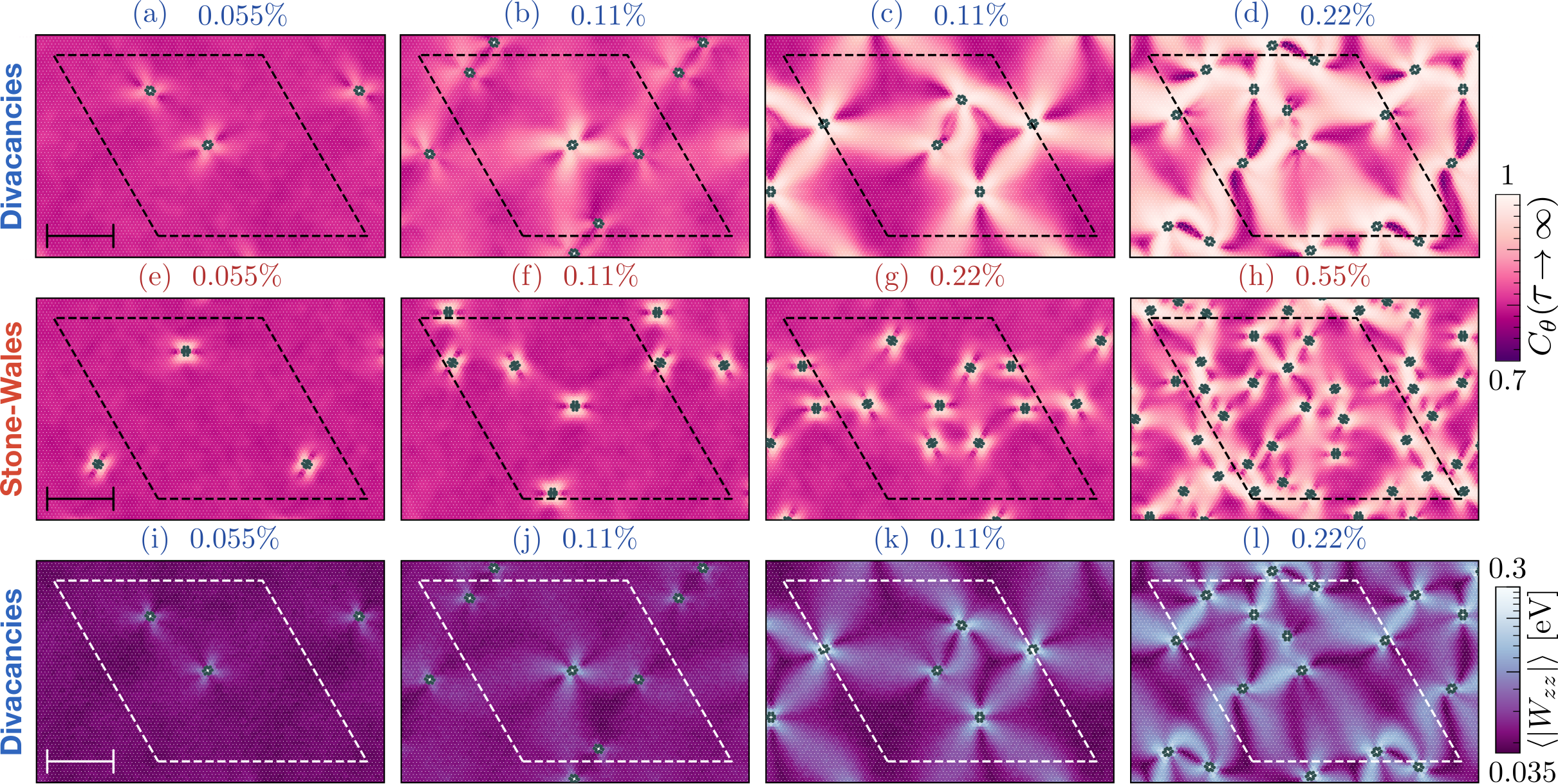}
    \caption{
    \cs{\textbf{Microscopic origin of the defect-induced dynamical transition.}
    Atomically resolved dynamics for (a-d) divacancies (DV) and (e-h) Stone-Wales (SW) defects, at increasing concentration from left to right. 
    The atoms are colored based on their long-time inclination correlation, except defects which are shown in green. Dark purple colors indicate mobile regions where the inclination decorrelates, while bright colors reveal static regions. Snapshots (i-l) correspond to the DV samples (a-d) where the atoms are colored based on the magnitude of the virial contribution to the normal stress $\langle |W_{zz}| \rangle$. The static regions are caused by elastic interactions between defects which emerge at high concentration. Simulation boxes are delineated with dashed lines, and scale bars represent 5 nm.}
    \label{fig:maps}}
\end{figure*}

To rationalise these observations, we resolve the rippling dynamics at the atomic scale. 
We show a selection of snapshots in Fig.~\ref{fig:maps}. 
For a comprehensive overview of results and full computational details, the reader is referred to the section \cs{S1.B} and \cs{S2.B} of the SI.
Each atom is color-coded based on its atomic long-time inclination correlation \cs{such that dynamic regions appear dark, while static ones are bright}. \cs{From left to right, the concentration of DV (a-d, first row) and SW (e-h, second row) defects increases.}
The snapshots reveal distinct local dynamics influenced by defects, their concentration and spatial arrangement.
\cs{At dilute concentrations, the inclination dynamics is clearly constrained in the vicinity of both types of defects, see Fig.~\ref{fig:maps}(a, e).
For divacancies, the dynamical influence of defects extends over a slightly larger lengthscales than SW defects, and gradually fades from 1-2 nanometers away from the DV site. Beyond, the defects have no impact on the dynamics and most atoms exhibit a pristine correlation $\approx 0.8$.} \\

\cs{Next, we focus on concentrations lying around the transition region for DV defects. 
We present in Fig.~\ref{fig:maps}(b, c) two systems at $0.11\%$ DV, which only differ in the defects' arrangement. 
We observe clear sample-to-sample fluctuations, with one system (b)  similar to the dilute case, and the other one (c) qualitatively different. 
Such a strong variance (b, c) is due to the finite size of the samples investigated, and suggest the existence of a genuine disorder-induced transition in the limit of large samples. 
Beyond the transition, see Fig.~\ref{fig:maps}(c, d), the defects have a system-spanning impact on the inclination dynamics. We observe the emergence of dynamically frozen paths connecting defects which can be lie more than 10 nm apart. This contrasts greatly with their local (1-2 nm) impact observed at low concentrations.} \\

\cs{These observations help us rationalise the DV results in Fig.~\ref{fig:quant_results}(c). At low concentration, the defects act as local pinning sites for the dynamics. In this regime, their effect is additive, hence a linear increase in the average $C_\theta(\tau \to \infty)$. Around 0.1\%, the defects start to interact over large lengthscales (in particular, larger than the interatomic potential cut-off), leading to the dynamical pinning of a large proportion of atoms lying between defects, and a sharp increase in the inclination correlation. We discuss below the nature of these defect-defect interactions.} \cs{Conversely, the Stone-Wales defects give rise to a qualitatively similar, but quantitatively different, scenario. We see in Fig.~\ref{fig:maps}(e, f, g) that the impact of SW defects remains short-ranged up to much larger concentrations than DV defects. This explains why the linear increase of $C_\theta$ extends up to large concentrations for the SW defects. At 0.22\%, the SW defects (g) have a local impact while the system is well beyond the transition for DV defects (d). Only at much larger concentrations, \textit{e.g.} 0.55\% (h) do SW defects give rise to dynamically frozen paths connecting them. A much smaller fraction of the atoms is affected by these defect-defect interactions, hence the absence of a sharp jump in the SW data of Fig.~\ref{fig:quant_results}(c) around these concentrations.} \\

\cs{We now turn to the mechanism underpinning this defect-induced dynamical transition. Atomic defects disrupt the hexagonal, minimal energy structure thereby generating stresses within the material. Indeed, the reconstruction of a DV defect, see inset of Fig.~\ref{fig:quant_results}(a), affects bond lengths in its vicinity. To reveal these defect-induced stresses, we compute the per-atom virial contribution to the stress tensor. More specifically, we focus on the normal contribution, which we denote $W_{zz}$. Given the $z\rightarrow -z$ symmetry, we report its absolute value, and average it over time to filter out thermal fluctuations. More details on this measurement are provided in the Methods. We report in Fig.~\ref{fig:maps}(i-l) the atomic normal stresses $\langle |W_{zz}| \rangle$ for the DV samples of panels (a-d). The similarity between the stress and dynamic maps is striking. This is true for SW defects too, for which the stress maps (see SI figure S17) are similar to the dynamic ones (e-h). Our analysis demonstrates that the dynamic-to-static transition emerges from elastically-mediated interactions between defects. The dynamically frozen paths correspond to regions characterized by large stresses. The dynamic regions are instead characterized by elastic stresses which can be overcome by thermal fluctuations since $k_B T\approx 0.03$ eV at room temperature, to be compared to the values reported in Fig.~\ref{fig:maps}(i-l), from 0.035 to 0.3 eV.}\\

\cs{Beyond the transition in DV samples, in addition to the dynamically frozen paths, we find regions extending over a few nanometers which exhibit $C_\theta(\tau \to \infty)$ values lower than those of the pristine reference, indicating more dynamic behaviour. In fact, the darkest (most mobile) regions are found at the highest defect concentration, Fig.~\ref{fig:maps}(d).} A careful investigation reveals that these areas undergo thermally-activated mirror buckling events during which their curvature flips. This phenomenon was first evidenced experimentally in pristine graphene \cite{Ackerman2016}, and phase-field models 
\cite{AlaNissila2022, AlaNissila2023} helped reveal that they emerge from global compression induced by boundary effects. \cs{Our elastic analysis shows that defects also generate stresses, and thus give rise to nanoscale mobile regions which undergo mirror-buckling events.} This could enable the creation of membranes with programmable memory~\cite{plummer2}. The presence of such local mirror-buckling regions at intermediate densities could help explain the experimental observation of a non-monotonic Young's modulus in defective graphene \cite{Lopez-Polin2015}.

\section{Discussion}

\begin{figure}[t!]
    \centering
    \includegraphics[width=0.35\textwidth]{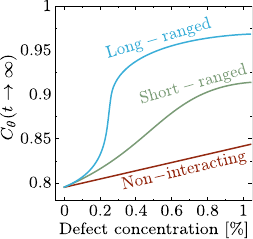}
    \caption{
    \textbf{Impact of Defect Interactions on Rippling Dynamics.} 
    This sketch illustrates how rippling dynamics is influenced by varying concentrations of generic defects. In the absence of defect interactions, defects affect the dynamics locally, leading to a linear increase in inclination correlation with defect concentration. As defect-defect interactions grow stronger and extend over longer ranges, the response transitions from a linear increase to an S-shaped curve, eventually culminating in a sharp transition. The axis correspond to typical defects in graphene, but could differ for other 2D materials.
    \label{fig:scenario}}
\end{figure}

\cs{We have investigated the impact of two defects on the rippling dynamics of free-standing graphene sheets. Our analysis reveals the existence of a sharp dynamic-to-static transition in the case of DV defects, which are found to strongly couple via the elasticity of the sheet. The SW defects, on the contrary, interact on much smaller lengthscales, and give rise to smaller stresses. Their impact on the dynamics is thus much smoother as their concentration increases. We summarize our results in a sketch Fig.~\ref{fig:scenario}, in which we extrapolate our findings to generic defects. Any atomic defect will give rise to local stresses, which in turn pin the dynamics locally. If these defects do not couple, their impact on the dynamics will be additive, yielding a linear increase in the inclination correlation (bottom curve). If the local stresses couple elastically, and the interactions are relatively short-ranged and weak, the curve will smoothly depart from a linear behaviour. Instead, if the defects give rise to large stresses which couple elastically over larger lengthscales, this will induce a sharp dynamic-to-static transition at a concentration inversely proportional to this interaction lengthscale (top curve).}
\cs{While stronger defects like dislocations and disclinations are known to cause larger stresses and surface corrugation~\cite{Liu2010}, our study demonstrates that even smaller lattice distortions can lead to a phase transition to static rippling.}
\cs{Importantly, the qualitative mechanism driving the transition does not depend on the material's precise chemical composition but instead on the generic elastic properties of two-dimensional sheets. 
Consequently, similar behavior is expected in defective hBN, TMDC monolayers, and other two-dimensional materials, given their structural similarity to graphene.}\\

\cs{Our results were obtained for free-standing graphene sheets. 
To explore the relevance of the observed dynamic-to-static transition in more realistic settings, we first consider experimental configurations where free-standing graphene is prepared by attaching its edges to a substrate, leaving the central portion suspended over a hole.
This attachment introduces additional stresses, which may combine with defect-induced internal stresses. 
However, the experimental observation of dynamic rippling in pristine samples suggests that these stresses remain small enough for the present phenomenon to occur in suspended monolayers. 
Notably, the critical defect concentration identified in our study aligns quantitatively with the maximum in the non-monotonic behavior of the Young's modulus \cite{Lopez-Polin2015}, indicating a strong link between rippling dynamics and mechanical properties \cite{Nicholl2015}. 
This connection is further supported by our stress analysis.}\\

\cs{In many cases, the monolayer is adsorbed onto a substrate. Van der Waals interactions will likely influence the phenomenon described here.
For graphene on metallic surfaces, the typical adsorption energy per atom is relatively small, in the meV range~\cite{WINTTERLIN20091841,PhysRevB.84.201401,PhysRevB.85.085425}. 
These interaction energies should be compared with the values reported in Fig.~\ref{fig:maps}(i-l). 
While the interaction with the substrate is expected to significantly reduce dynamic rippling in pristine samples, they could be weak enough for the static corrugated phase to persist.
Our study highlights the potential to leverage atomic defects to engineer nanoscale regions with pronounced curvature in graphene sheets. 
Periodically corrugated graphene and hexagonal boron nitride (hBN) can exhibit strong pseudomagnetic fields~\cite{levy2010} and exotic strongly correlated electronic states~\cite{mao2020evidence,gupta2022designing}. 
These corrugated structures arise from adsorption on a flat substrate with different thermal properties or directly on a corrugated substrate. 
We have investigated disordered arrangements of defects, but preliminary results show that ordered arrays of defects can generate similar periodically corrugated free-standing 2D materials. 
Such defect-engineered materials thus have the potential to exhibit unique and intriguing physical behaviors.}\\

\cs{The scenario of defective graphene immersed in liquids presents particularly interesting applications. For example, water has been shown to stabilize vacancy defects in graphene~\cite{Sharma_2019}. At the static level, defects are known to alter the wetting behavior of graphene~\cite{Li2013}.
When used as a confining material in nanofluidic flows, the static-to-dynamic transition in graphene can significantly affect fluid flow along the membrane, either enhancing or inhibiting it~\cite{marbach_transport_2018}. 
It would be particularly interesting to extend defect engineering studies of solid friction~\cite{Zambudio2021} to liquid-solid friction~\cite{Joly2016}.
While the fate of the dynamic-to-static transition in defective graphene when fully immersed in water remains open, Ref.~\cite{Deng2022} showed that aligned divacancy defects constrain the diffusion of nano droplets across graphene sheets. 
Such behavior could stem from the static transition discussed here, highlighting the critical role of these defects in modulating transport properties.
The static phase, where rippling dynamics are suppressed, makes defective graphene a promising candidate for applications such as DNA sequencing. 
In this context, graphene rippling dynamics induces significant noise that limits the measurements of DNA translocation through nanopores~\cite{Heerema_2015}. 
The defect-induced static transition evidenced here could reduce such noise and enable efficient DNA sequencing. }

\section{Conclusion}

We have reported a large-scale machine learning-driven molecular dynamics study on the impact of atomic defects on the rippling dynamics in graphene.
We introduce a novel approach based on the local inclination to quantify both corrugation and time-dependent fluctuations, which can be compared to experimental electron diffraction patterns.
We unveil a disorder-driven dynamical transition from freely propagating ripples to frozen rippling dynamics upon increasing defect concentration. 
\cs{We demonstrate that this transition is driven by elastically-mediated interactions between defects. We discuss how different defects, and hence elastic interactions, influence the nature of this transition.}\\

Obtaining direct experimental confirmation of the predictions reported here, \textit{e.g.} with ultrafast electron crystallography \cite{Hu2016}, will make interesting work for the future. 
On the theoretical side, future work based on discrete real space models and nonlinear continuum elastic models will be required to fully characterize the disorder-driven transition predicted here.
In particular, past works \cite{plummer1,plummer2,plummer3} on thin elastic sheets comprising ordered dilation impurities revealed interesting critical phenomena and physical behaviour. Extending these works to randomly positioned compression impurites, mimicking the atomic defects, is of great interest. 
Our analysis of local sheet fluctuations provides initial mechanistic insights, highlighting the pivotal role of defect interactions in implementing pathways or areas of desired mobility.\\

Overall, our work attains an unprecedented atomic-level resolution of graphene dynamics, with findings readily applicable to the broad spectrum of 2D materials.
This offers exciting prospects for the development of innovative devices, in which the behavior of ripples, motion of adsorbates, and the sheet's catalytic activity are directly controlled through reflection and refraction at defects. 
Our work demonstrates the potential of employing atomic defects towards various applications and the possibility of tailored directional phonon-mediated flow in nanofluidics. 

\section{Methods}

\subsection{\cs{Setup of defective samples}}

The pristine graphene samples are composed of 7200 carbon atoms.
The defective ones are created starting from a perfectly flat and pristine graphene sheet, and iteratively removing two adjacent carbon atoms. 
The reconstruction of divacancy defects is not enforced in the initial configuration and occurs spontaneously during the simulation. We do not observe any defect migration.
The position and orientation of the divacancies are chosen randomly, yet we ensure a minimum separation of 10 $\mathrm{\AA}$ between defect centers. 
Following previous works \cite{Mortazavi2013,Thiemann2021}, we define the defect concentration as the ratio of removed atoms to the total number of atoms in a pristine graphene sheet.
We study defect concentrations from $\approx 0.03\%$ (corresponding to an isolated defect) to a high level of $1\%$ (corresponding to 36 divacancy defects, or 7128 carbon atoms).
We show representative snapshots of pristine and highly defective samples in Fig.~\ref{fig:overview}.
For each defect concentration, we run several simulations placing the divacancies at various positions and orientations.

\subsection{Molecular Dynamics Simulations}

We use the machine learning-based Gaussian approximation potential~\cite{Bartok2010a} (GAP) for carbon, GAP-20 \cite{Rowe2020b}. 
\cs{By decomposing atomic interactions into 2-body, 3-body, and many-body (SOAP) contributions, this model achieves a balance between short-range and long-range interactions, making it ideal for applications like this study.}
Unlike simpler classical force fields for carbon, the GAP-20 potential has been shown to reliably predict a variety of properties of graphene, including its phonon dispersion spectrum, elastic constants, defect structure and formation energies \cite{Rowe2020b}, and provides first principles quality even at high defect concentrations \cite{Thiemann2021}.
\cs{Additionally, Section S3.A.1 of the SI provides further evidence that GAP-20 accurately reproduces the DFT-predicted interaction energy of divacancies across various separations.}
A detailed description of the GAP framework can be found elsewhere~\cite{Bartok2010a,Bartok2015,Deringer2021}. \\

The simulations were performed at 300 K and zero stress, with periodic boundary conditions, using the LAMMPS software package \cite{Plimpton1997}. A detailed description of the simulation setup can be found in section S2.A of the SI.
The entire post-processing analysis for all simulations was performed in Python using the ASE \cite{HjorthLarsen2017}, MDAnalysis \cite{MichaudAgrawal2011,Gowers2016}, and OVITO \cite{Stukowski2010} software packages.
Computational details and additional convergence tests for both system size and simulation time are presented in sections S2.B and S4.A as well as S4.B of the SI, respectively.

\subsection{Measuring Corrugation and Dynamics}

We quantify the static corrugation of a graphene sheet based on the rms inclination which represents the average deviation of the surface normal from its mean ($z$) direction. 
For each atom, we compute the vector normal to the sheet by locally fitting the surface defined by neighboring carbon atoms (see section S2.B of the SI for details).
The normal inclination $\theta$, illustrated in Fig.~\ref{fig:overview}(g), is defined as the angle between the normal vector and the $z$ direction.
This approach extends the concept of the pyramidalisation angle\cite{pyramidalization} beyond the nearest neighbors and allows for a smooth and accurate representation of the local surface.
Our method fully accounts for the corrugated nature of the sheet and
allows a comparison with experimental efforts to measure graphene's corrugation via TEM diffraction patterns~\cite{Meyer2007}.\\

Notably, both experiments \cite{Singh2022} and computational work \cite{Singh2013} suggest an exponential relationship, $\theta_{\mathrm{rms}} \propto \exp{\left( - l/L \right)}$, where $L$ is the size of the freestanding graphene sheet and the parameter $l$ can be interpreted as the inverse resolution of the measurement.
Specifically, $l$ corresponds to the electron coherence length in TEM experiments, while in simulations, $l$ measures the patch size over which the corrugation is averaged.
In our approach, individually fitting the angle for each atom without any spatial averaging results in a basically infinite resolution of $l=0$.
This allows a direct comparison to electron diffraction experiments for graphene sheets of about 1~$\mu$m, where typical electron coherence lengths result in very low $l/L \leq 0.02$.
For a more details discussion on this aspect, the reader is referred to section S4.A.1 of the SI.\\

To assess the dynamics of the graphene sheet, we compute the autocorrelation function of the normal inclination, normalised by its equal-time value, which we denote $C_\theta(\tau)$.
This function tracks the variation of the surface normal over time, providing valuable insights into the system's dynamic behaviour.
To enable a straightforward comparison across systems, we fit an exponential function to $C_\theta(\tau)$ and extract the plateau value, denoted as $C_\theta(\tau \to \infty)$.
Comprehensive details are provided in section S2.B of the SI.\\

\subsection{\cs{Normal stress}}
\cs{In order to assess the elastic state of the sheet at varying defect concentrations, we measure a quantity related to the normal stress. 
Specifically, we focus on the virial contribution of each atom to the out-of-plane stress, $W_{zz}$.
This property is computed directly in LAMMPS \cite{Plimpton1997} using the approach introduced by Thompson \textit{et al.} \cite{thompson2009} for many-body potentials.
 Given that the per-atom values are noisy, we compute averages over a trajectory of 1 ns (1000 configurations separated by 1 ps). The time-averaged data $\langle | W_{zz}| \rangle$ are shown in the main text figure 3, where the absolute values are taken to account for the $z \rightarrow -z $ symmetry.}

\section*{Data and code availability}
The code to compute the normal inclination and its autocorrelation function is open source and can be accessed free of charge under \href{https://github.com/flt17/graphene-analysis}{https://github.com/flt17/graphene-analysis}.
The MD trajectories of the systems depicted in Figure 3, along with the LAMMPS input files, are available at \href{https://zenodo.org/records/14697820}{https://zenodo.org/records/14697820}.

\section*{Supplementary Information}

See the Supporting Information for links to movies of the temporal evolution over 0.1~ns of different graphene sheets including the pristine reference as well as the four \cs{divacancy} systems depicted in Fig.~\ref{fig:maps}, a detailed description of the molecular dynamics simulations (S1.A), the computation of the normal inclination and its autocorrelation function (S1.B), a comprehensive overview of our results for all systems (S2), an analysis of \cs{the performance of the GAP-20 to capture the interaction between defects and investigation} of the dependence of our results on system size and simulation duration (S3.A), and comparison with alternative observables (S3.B).

\section*{Acknowledgements}

The authors thank Lydéric and Marie-Laure Bocquet, Mathieu Lizée, Nicolas Romeo and Vincenzo Vitelli for fruitful discussions. 
We are grateful to the UK Materials and Molecular Modelling Hub for computational resources, which is partially funded by EPSRC (EP/P020194/1 and EP/T022213/1).
Through our membership of the UK’s HEC Materials Chemistry Consortium, which is funded by EPSRC (EP/L000202 and EP/R029431), this work used the ARCHER and ARCHER2 UK National Supercomputing Service.
We are also grateful for the computational resources granted by the UCL Grace High Performance Computing Facility (Grace@UCL), and associated support services. 
C.~S.~acknowledges support from a Herchel Smith Fellowship, University of Cambridge and a Ramon Jenkins Research Fellowship from Sidney Sussex College, Cambridge in the early stages of the project, as well as a CNRS Physique ``Tremplin'' funding.
A.~M.~acknowledges support from the European Union under the “n-AQUA” European Research Council project (Grant No. 101071937).

\section*{Author contributions}

F.L.T., C.S., E.A.M., A.M. designed the research, F.L.T., C.S. performed the research, F.L.T. performed the molecular dynamics simulations and developed the data analysis code, F.L.T., C.S. analysed the data, F.L.T., C.S., E.A.M., A.M. wrote the paper, E.A.M., A.M. provided supervision and funding.

%----------------------------------------------------------------------
\bibliography{revision}
%----------------------------------------------------------------------

\end{document}